\documentclass[11pt]{article}
\usepackage{amsmath}
\usepackage[dvips]{graphicx}
\begin{document}
\title{Failure of energy stability in Oldroyd-B fluids
at arbitrarily low Reynolds numbers}

\author{Charles R. Doering$^{1,2}$, Bruno Eckhardt$^{3}$ and
J\"org Schumacher$^{3,4}$
\\
{$^{1}$\emph{Department of Mathematics, University of Michigan,
Ann Arbor, MI 48109-1043 USA}}
\\
{$^{2}$\emph{Michigan Center for Theoretical Physics,
Ann Arbor, MI 48109-1040 USA}}
\\ 
{$^{3}$\emph{Fachbereich Physik, Philipps-Universit\"at Marburg, 
35032 Marburg,
Germany}}
\\
{$^{4}$Present address: \emph{Department of Mechanical Engineering,}}
\\
\emph{Technische Universi\"at Ilmenau, 98684 Ilmenau, Germany}
\\
}
\maketitle 

\begin{abstract}
Energy theory for  incompressible Newtonian fluids is, in many cases, capable of 
producing strong absolute stability criteria for steady flows.
In those fluids the kinetic energy naturally defines a norm in which perturbations 
decay monotonically in time at sufficiently low (but non-zero) Reynolds numbers.    
There are, however, at least two obstructions to the generalization of such 
methods to Oldroyd-B fluids.
One previously recognized problem is the fact that the natural energy does not 
correspond to a proper functional norm on perturbations.
Another problem, original to this work, is the fact that fluctuations in Oldroyd-B 
fluids may be subject to non-normal amplification at arbitrarily low Reynolds 
numbers (albeit at sufficiently large Weissenberg numbers). 
Such transient growth, occuring even when the base flow is linearly stable, 
precludes the uniform monotonic decay of any reasonable measure of the 
disturbance's amplitude.\\
\\
{\bf Keywords:} Energy stability, planar shear flow, viscoelastic flow
\end{abstract}

\section{Introduction}

Linear stability theory can predict sufficient conditions for instability, but some sort of
nonlinear analysis is necessary to determine stability against finite amplitude perturbations.
In many fluid dynamic systems the notion of nonlinear energy stability is useful 
for determining conditions for absolute stability \cite{Joseph,Straughan}.
Energy stability theory exploits the observation that physically natural 
energy functionals may generate norms that can be used to measure the amplitude of 
disturbances about a base flow.
Energy analysis typically leads to sufficient conditions for the norms of arbitrary amplitude
disturbances to decay monotonically in time, a very strong notion of absolute stability.
For example, steady states of incompressible Newtonian fluids are very generally energy
stable at sufficiently low Reynolds numbers.
In some cases, notably simple Rayleigh-Benard convection problems in the Boussinesq
approximation, the critical Rayleigh numbers for energy stability and linear instability coincide
to establish the supercritical nature of the bifurcation point \cite{DR}.
Moreover, energy methods can be generalized to derive rigorous bounds on flow quantities
for high Reynolds (or Rayleigh) number steady, unsteady and even turbulent flows
\cite{Howard,Busse,DC}.

Elastic and kinetic energy budgets have been used frequently in the past for
the linear stability analysis of dilute polymer solutions in simple flow geometries.
Purely elastic instabilities at small Reynolds numbers were analyzed by Joo and Shaqfeh
\cite{Joo1} (a review can be found in \cite{Shaqfeh96}) 
for an Oldroyd-B fluid in a Taylor-Couette geometry. 
In the total disturbance energy balance, coupling between perturbations of the 
velocity field and gradients of the polymer stress showed the same temporal 
growth as the kinetic energy of flow and were thus identified as the ones causing 
the instability.  
Similar energy budget considerations were applied and extended by Sureshkumar for 
linear stability analysis of plane Poiseuille flows at larger Reynolds number 
\cite{Sureshkumar01} where the importance of this coupling was confirmed. 
For growing Weissenberg numbers the coupling of velocity perturbations to 
gradients of normal viscoelastic stresses was found to dominate the 
destabilization.
 
In this paper we investigate the notion of nonlinear energy stability for 
Oldroyd-B fluids.
To set the stage and fix notation, first recall the case of plane Couette flow of 
a Newtonian fluid.
The fluid is confined between parallel horizontal plates at $y=0$ and $y=\ell$, 
with the lower plate at rest and the upper plate moving in the $x$-direction 
(${\bf i}$) with velocity $U$.
For simplicity, we consider periodic conditions in the streamwise (${\bf i}$) and 
cross-stream (${\bf k}$) directions.
Measuring length in units of $\ell$ and time in units of $\ell/U$, the velocity 
vector field ${\bf u}({\bf x},t) = \textbf{i}u +\textbf{j}v+\textbf{k}w$ of an 
incompressible Newtonian fluid satisfies the Navier-Stokes equations
\begin{equation}
\frac{\partial {\bf u}}{\partial t} + {\bf u}\cdot{\bf \nabla}{\bf u}+
{\bf \nabla}p = \frac{1}{Re} \Delta {\bf u}, \ \ \ \ \
{\bf \nabla}\cdot{\bf u} = 0,
\end{equation}
where $Re=U\ell/\nu$ is the Reynolds number defined in terms of the kinematic viscosity of
the fluid $\nu$.

The kinetic energy evolution equation is
\begin{equation}
\frac{d}{dt} \frac{1}{2} \int{|\textbf{u}({\bf x},t)|^{2}dx dy dz} =
- \int{\frac{1}{Re}|\nabla \textbf{u}|^{2} dx dy dz} 
+\int\frac{1}{Re} \frac{\partial u}{\partial y}\Big|_{y=1} dx dz.
\end{equation}
The negative definite term on the right hand side is the bulk viscous energy dissipation rate,
and the indefinite term is the power expended by the agent enforcing the boundary condition.

Plane Couette flow, an exact steady solution in this geometry at every Reynolds number, is
\begin{equation}
{\bf u}_{Couette} = \textbf{i}y.
\end{equation}
If we perturb plane Couette flow at time $t=0$ with an initial condition
\begin{equation}
{\bf u}({\bf x},0) = \textbf{i}y + \delta\textbf{u}({\bf x},0),
\end{equation}
then the energy in the subsequent disturbance $\delta\textbf{u}({\bf x},t)$ evolves according to
\begin{equation}
\frac{d}{dt} \frac{1}{2}\int{|\delta \textbf{u}({\bf x},t)|^{2} dx dy dz} =
- \int{(\frac{1}{Re}|\nabla \delta \textbf{u}|^{2}+\delta u \delta v) dx dy dz}.
\end{equation}
Energy stability theory ensures us that there is a positive critical value $Re_{ES} \approx 83$ \cite{Joseph,Henningson}
such that $Re < Re_{ES}$ guarantees that there exists a number $\alpha > 0$ such that
\begin{equation}
\int{(\frac{1}{Re}|\nabla \delta \textbf{u}|^{2}+\delta u \delta v) dx dy dz}
>  \frac{\alpha}{2}\int{|\delta \textbf{u}|^{2} dx dy dz}.
\end{equation}
This in turn implies
\begin{equation}
\int{|\delta \textbf{u}({\bf x},t)|^{2} dx dy dz}
< \int{|\delta \textbf{u}({\bf x},0)|^{2} dx dy dz}
\times e^{-\alpha t}
\end{equation}
so the norm of the perturbation decays monotonically (and exponentially) as 
$t \rightarrow \infty$.

This is the sense in which plane Couette flow is absolutely stable to arbitrary (finite energy)
disturbances when $Re < Re_{ES}$.
Plane Couette flow is linearly stable at all $Re$ \cite{Romanov}, but finite amplitude
disturbances may not decay sufficiently high above the energy stability limit.
Non-normal amplification of small disturbances, a transient growth phenomenon available even in
linearly stable systems, is recognized as an effective mechanism for stimulating finite-amplitude
instabilities \cite{Henningson,grossmann2000}.

In the following sections we address the question of whether or not this notion of 
energy stability may be fruitfully generalized to Oldroyd-B fluids.
There have been previous analyses in this direction for other non-Newtonian 
constitutive relations \cite{Preziosi}, and {\it a priori} energy estimates for 
Oldroyd-B fluids \cite{Owens}, but to our knowledge there has been no definitive 
determination of whether an energy stability principle exists for Oldroyd-B 
fluids.
Here we show by example that no such principle is possible in general for 
Oldroyd-B fluids.
By examining particular exact solutions to the full nonlinear equations of motion
we determine that some perturbations will grow, absorbing energy from the base 
flow, even at arbitrarily small (but nonvanishing) Reynolds numbers.

\section{Energy evolution for Oldroyd-B fluids in the Couette geometry}

In nondimensional form, the equations of motion for an Oldroyd-B fluid in the plane
Couette geometry\footnote{That is, for fluid confined between parallel horizontal plates
at $y=0$ and $y=1$ with the lower plate at rest and the upper plate moving in the
$x$-direction (${\bf i}$) with velocity $1$ and, for simplicity, periodic conditions in
the streamwise (${\bf i}$) and cross-stream (${\bf k}$) directions.} are
\begin{equation}
\frac{\partial {\bf u}}{\partial t} + {\bf u}\cdot{\bf \nabla}{\bf u} +
{\bf \nabla}p = \frac{1}{Re} \Delta {\bf u} + {\bf \nabla} \cdot {\bf \tau},
\ \ \ \ \ {\bf \nabla}\cdot{\bf u} = 0,
\label{OB1}
\end{equation}
where $Re=U\ell/\nu$ is the Reynolds number.
The polymer stress tensor ${\bf \tau}$ is
\begin{equation}
{\bf \tau} = \frac{s}{Re Wi}(\textbf{c}-\textbf{I})
\label{OB2}
\end{equation}
where the Weissenberg number $Wi = U\lambda/\ell$ is the product of the polymer
relaxation time $\lambda$ and the imposed rate of strain $U/\ell$, the parameter $s$ is the coupling
constant proportional to the concentration of the polymers in the flow, and the polymer
configuration tensor $\textbf{c}(\textbf{x},t)$ evolves according to
\begin{equation}
\frac{\partial {\bf c}}{\partial t} + {\bf u}\cdot{\bf \nabla}{\bf c} =
\textbf{c}\cdot{\bf \nabla}\textbf{u} + ({\bf \nabla}\textbf{u})^{T}
\cdot\textbf{c} + \frac{1}{Wi}(\textbf{I}-\textbf{c}).
\label{OB3}
\end{equation}
The component $c_{ij}$ of the configuration tensor $\textbf{c}$ is the
end-to-end moment $\left< r_{i} r_{j} \right>$ of the extension of the polymers,
modeled as linear springs.
In this scaling, the equilibrium configuration corresponds to an isotropic distribution with unit
end-to-end displacements.
The elastic energy (density) stored in the polymers is proportional to
$\frac{1}{2} tr \textbf{c} \sim \frac{1}{2} tr({\bf \tau}) + constant$.

The kinetic energy evolution equation is
\begin{equation}
\frac{d}{dt} \frac{1}{2} \int{|\textbf{u}({\bf x},t)|^{2} dx dy dz} =
- \int{\frac{1}{Re}|\nabla \textbf{u}|^{2} dx dy dz} 
+\int_{y=1}(\frac{1}{Re} \frac{\partial u}{\partial y} + {\bf \tau}_{12}) dx dz
- \int{{\bf \tau}:\nabla \textbf{u} dx dy dz}.
\end{equation}
The first (negative definite) term on the right hand side is the bulk viscous energy dissipation rate;
the second (indefinite) term is the power expended by the agent enforcing the boundary condition;
the third (indefinite) term corresponds to the rate of work done by the polymers 
on the flow, an energy exchange term.
The elastic energy stored in the polymers evolves according to
\begin{equation}
\frac{d}{dt} \frac{1}{2} \int{tr({\bf \tau}) dx dy dz} =
\int{{\bf \tau}:\nabla \textbf{u} dx dy dz} 
- \frac{1}{Wi}\int{tr({\bf \tau}) dx dy dz}.
\end{equation}
The two terms on the right hand side are, respectively, the rate of work done by the flow on the
polymers and the dissipation rate due to polymer relaxation processes.
Hence the total energy is identified as the sum of the fluid's kinetic energy and the polymers'
elastic potential energy,
\begin{equation}
{\cal E}(t) = \int{\frac{1}{2}[|\textbf{u}({\bf x},t)|^{2}
+ tr({\bf \tau})] dx dy dz},
\label{EOB}
\end{equation}
and the total energy evolution is
\begin{equation}
\frac{d{\cal E}}{dt} = 
- \int{\frac{1}{Re}|\nabla \textbf{u}|^{2} dx dy dz} 
- \frac{1}{Wi}\int{tr({\bf \tau}) dx dy dz}
+\int_{y=1}(\frac{1}{Re} \frac{\partial u}{\partial y}
+ {\bf \tau}_{12}) dx dz,
\label{energyevolution}
\end{equation}
expressing the net balance between dissipation (the first two negative terms on the right hand
side) and the work done on the fluid by the agency imposing the boundary condition (the 
indefinite surface integral of the stresses).

Two distinct obstructions stand in the way of attempts to generalize the property 
of energy stability to Oldroyd-B fluids.
The first problem is evident already in (\ref{EOB}): while the kinetic energy 
defines a natural norm on the velocity field, the polymer energy does not 
correspond to a norm (or even a metric) on the polymer stress field \cite{Owens}.
Indeed, the physically relevant space for the polymer configuration tensor 
$\textbf{c}$ is the set of positive symmetric matrices, but this does not 
constitute a linear vector space.
So strictly speaking the concept of a norm is not meaningful.
More importantly, though, while it is true that for positive symmetric matrices 
$tr(\textbf{c})=0$ implies $c=0$, the trace does not satisfy a triangle 
inequality necessary for a metric.
It is clear, then, that the most straightforward generalization of energy 
stability will not work for Oldroyd-B fluids.

Nevertheless, various norms or metrics can be defined on perturbations
($\delta\textbf{u}$, $\delta \tau$) of the variables.
For example, an $L^{2}$ norm could be defined by
\begin{equation}
\|\delta\textbf{u},\delta \tau\|^{2} =
\int{[\delta \textbf{u}_{i}\delta \textbf{u}_{i} +
\delta \tau_{ij}\delta \tau_{ij}] dx dy dz},
\end{equation}
Or a functional like
\begin{equation}
{\cal M}(\delta\textbf{u},\delta \tau)=
\int{[\delta \textbf{u}_{i}\delta \textbf{u}_{i} +
\sqrt{\delta \tau_{ij}\delta \tau_{ij}}] dx dy dz}
\end{equation}
could be used as an effective measure of the amplitude of perturbations.
In any case there is a second fundamental---in this case 
dynamical---obstruction to a principle of energy stability in Oldroyd-B 
fluids:
As will be shown in the next section, the problem is that it is possible 
for perturbations to absorb energy from a base flow and, by any reasonable 
measure, grow at arbitrarily low $Re$.
That is, it is {\it not} possible to assert in any generality that there 
is an analog of energy stability in the sense that there is a {\it 
positive} critical value $Re_{ES}$ so that $Re < Re_{ES}$ implies that all 
perturbations decay monotonically.
This ``no-go" result is established by producing specific examples where 
disturbances do not decay monotonically at arbitrarily low $Re$.

\section{Some exact solutions in the Couette geometry}

Plane Couette flow $\textbf{u}_{Couette} = \textbf{i}y$ is an exact solution of 
the Oldroyd-B equations (\ref{OB1}-\ref{OB3}) in the plane shear geometry when
augmented with the constant (in space and time) polymer stress tensor
\begin{equation}
\textbf{c}_{Couette} =
\begin{pmatrix}
1+2Wi^{2} & Wi & 0 \\
Wi & 1 & 0 \\
0 & 0 & 1
\end{pmatrix} .
\end{equation}
This flow is linearly stable for upper convected Maxwell fluids \cite{Renardy}, i.e.,
when $s = {\cal O}(Re) \rightarrow \infty$.
Consider perturbations with $\delta \textbf{u}(\textbf{x},t) = 0$ and
\begin{equation}
\delta \textbf{c}(t)=
\begin{pmatrix}
\delta c_{11}(t) & \delta c_{12}(t) & 0 \\
\delta c_{21}(t) & \delta c_{22}(t) & 0 \\
0 & 0 & 0
\end{pmatrix}
\end{equation}
where, preserving the symmetry, $\delta c_{12} = \delta c_{21}$.
That is, we look for solutions where the polymer stress tensor is a function of time, but not space,
so there is no influence of the polymers back onto the flow field.
For this class of solutions there are no perturbations of
the velocity field or the Newtonian stress.

Inserting this ansatz into the full nonlinear equations of motion produces
an exact closed set of {\it linear} ordinary differential equations for
the components of $\delta \textbf{c}$:
\begin{align}
\frac{d \delta c_{11}}{dt} &= -\frac{1}{Wi} \delta c_{11} + 2 \delta c_{12}, \\
\frac{d \delta c_{12}}{dt} &= -\frac{1}{Wi} \delta c_{12} + \delta c_{22}, \\
\frac{d \delta c_{22}}{dt} &= -\frac{1}{Wi} \delta c_{22}.
\end{align}
The general solutions, uniform in $Re$, are
\begin{align}
\delta c_{11}(t) &= [\delta c_{11}(0) + 2 \delta c_{12}(0)t + \delta c_{22}(0) t^{2}]
e^{-t/Wi},\\
\delta c_{12}(t) &=  [\delta c_{12}(0) + \delta c_{22}(0)t]
e^{-t/Wi},\\
\delta c_{22}(t) &= \delta c_{22}(0)e^{-t/Wi}.
\end{align}

These solutions display non-normal transient growth for a variety of
initial data as long as $Wi$ is not too small.
For example consider the zero energy (i.e, trace $\delta {\bf \tau}(0)=0$) perturbation
\begin{equation}
\delta \textbf{c}(0) =
\begin{pmatrix}
-a & 0 & 0 \\
0 & a & 0 \\
0 & 0 & 0
\end{pmatrix}
\end{equation}
where $0<a<1$.
The time-dependent perturbation of the conformation tensor is then
\begin{equation}
\delta \textbf{c}(t) = a e^{-t/Wi}
\begin{pmatrix}
t^{2}-1 & t & 0 \\
t & 1 & 0 \\
0 & 0 & 0
\end{pmatrix}.
\end{equation}
While the perturbations eventually decay exponentially (with algebraic 
modification), the transient dynamics can by highly non-monotonic.
For example the polymer-induced stress $\sim c_{12}(t)$ overshoots its steady 
value before relaxing.
The energy perturbation in the polymers for this solution is
\begin{equation}
\frac{\delta {\cal E}(t)}{{\cal E}_{steady-state}} = 
\frac{tr{\delta\textbf{c}(t)}}{3+2Wi^{2}} =
\frac{a t^{2} e^{-t/Wi}}{3+2Wi^{2}}.
\end{equation}
This disturbance initially grows nearly quadratically in time to a maximum at $t=2Wi$ where 
\begin{equation}
\frac{\delta {\cal E}}{{\cal E}_{steady-state}}{\big |}_{t=2Wi} = 
\frac{4a}{e^{2}}\frac{Wi^{2}}{3+2Wi^{2}},
\end{equation}
constituting an ${\cal O}(1)$ variation in polymer energy before decay sets in.
More significantly for our purposes is the evolution of a norm-like measure of the perturbation
such as the amplitude
\begin{equation}
A(t)=\frac{\sqrt{\delta c_{ij}(t)\delta c_{ij}(t)}}{\sqrt{\delta c_{ij}(0)\delta c_{ij}(0)}}
= e^{-t/Wi} \sqrt{1+\frac{t^{4}}{2}}.
\label{amplitude}
\end{equation}
Although $A(t)$ initially decreases, it displays a period of increase when
$Wi > \frac{4}{6^{3/4}} \approx 1.043$.
This behavior is plotted for several values of $Wi$ in Figure 1.

We have found non-normal features in the solutions for any $Wi > 1$.
Consider, for example, initial perturbations with
\begin{equation}
\delta \textbf{c}(0) =
\begin{pmatrix}
0 & a & 0 \\
a & 0 & 0 \\
0 & 0 & 0
\end{pmatrix}
\end{equation}
with $|a| < \sqrt{1+2Wi^{2}}-Wi$ to maintain the positive definiteness of \textbf{c}.
The amplitude of this disturbance evolves according to
\begin{equation}
A(t)=\frac{\sqrt{\delta c_{ij}(t)\delta c_{ij}(t)}}{\sqrt{\delta c_{ij}(0)\delta c_{ij}(0)}}
= e^{-t/Wi} \sqrt{1+2t^{2}},
\end{equation}
and it is easy to confirm that this amplitude exhibits transient
amplification so long as $Wi > 1$.

A particular example that gives some insight into a physical mechanism for the transient
growth behavior is the zero energy initial perturbation
\begin{equation}
\delta \textbf{c}(0) =
\begin{pmatrix}
-a & a Wi & 0 \\
a Wi & a & 0 \\
0 & 0 & 0
\end{pmatrix}
\end{equation}
where $|a| <<1$ corresponds to a (small) uniform rotation of the polymers from their
stationary configurations.
As illustrated in Figure 2, the initially rotated ellipsoid representing the mean 
polymer extension is stretched by the background shear well beyond the initial 
value---by an ${\cal O}(Wi^{2})$ factor at large $Wi$---before reassuming the 
stationary configuration.
During the period of transient growth, the polymers absorb energy from the 
background shear flow faster than the elastic relaxation can dissipate it.

\section{Summary and discussion}

The conventional notion of energy stability of a stationary flow implies monotonic decay of some
meaningful measure of the magnitude of arbitrary amplitude disturbances.
The exact solutions developed in the preceeding section show that it is generally \textit{not}
possible to establish monotonic decay of arbitrary perturbations of steady flows of Oldroyd-B
fluids at finite Reynolds numbers.
These examples require a nonvanishing (and not arbitrarily small) Weissenberg number so there must
be some nonvanishing shear in the base flow, but the observed transient amplification of perturbations,
up to ${\cal O}(Wi^{2})$ in magnitude, are uniform in the fluid's kinematic viscosity and $Re$. 
While a base flow may be linearly stable, the `extra' degrees of freedom in the polymers may allow
for exact solutions displaying non-normal transient growth.

%These particular solutions, where the flow remains steady but the polymers deform and evolve,
%are reminiscent of the recent experiments by Hur {\em et al} \cite{Shaqfeh}.
%The associated stress overshoots for $Wi > {\cal O}(20)$ in \cite{Shaqfeh} are similar in nature to
%those in the solutions presented here for $Wi > 1$.
%The initial conditions and constitutive relations are not exactly the same, but this correspondence suggests
%that similar phenomena could possibly be observed experimentally in simple Oldroyd-B fluids by measuring
%the transient stress response to a rapidly switched shear rate at low $Re$.

We remark that the problems extending energy stability techniques pointed out here are not
mitigated by using another physically motivated energy functional where a thermodynamic free energy
takes the place of the mechanical elastic energy alone \cite{EdwardsBeris}.  
That is, a free energy density of the form
\begin{equation}
{\cal F}(c) \sim tr(c) - \log{\det{c}}
\end{equation}
includes entropic effects so that the free energy minimum is precisely the equilibrium
polymer configuration with $c_{ij} = \delta_{ij}$.
However this still does not induce a proper metric or norm on configurations and because it does
not affect the equations of motion, the non-normal growth phenomena and the obstructions
it presents are still present.

Finally, after this work was completed we became aware of 
the analysis of Morozov and van Saarloos \cite{mvs} 
on % who demonstrate, by explicit calculation, 
an elastic finite-amplitude instability for sufficiently
high Weissenberg number and arbitrarily low Reynolds number.
That work goes beyond the observations presented here
by explicitly displaying an additional stationary
solution to the fully nonlinear 3-d equations.

\noindent
{\bf Acknowledgements:} 
We thank Dan Joseph, Joel Miller, Marcel Oliver, Anshuman Roy and Jean-Luc 
Thiffeault for helpful discussions.
We also acknowledge the stimulating remarks and suggestions of the referees.
One of us (CRD) acknowledges the hospitality of the Geophysical Fluid Dynamics 
Program at Woods Hole Oceanographic Institution where part of this work was 
completed.  
This research was supported in part by NSF Award PHY-0244859, an
Alexander von Humboldt Research Award (CRD), the PROCOPE 
program of the German Academic Exchange Service (JS), and the 
Deutsche Forschungsgemeinschaft.

\bibliographystyle{unsrt}

\clearpage

\begin{figure}
\centerline{\includegraphics[angle=0,scale=0.5,draft=false]{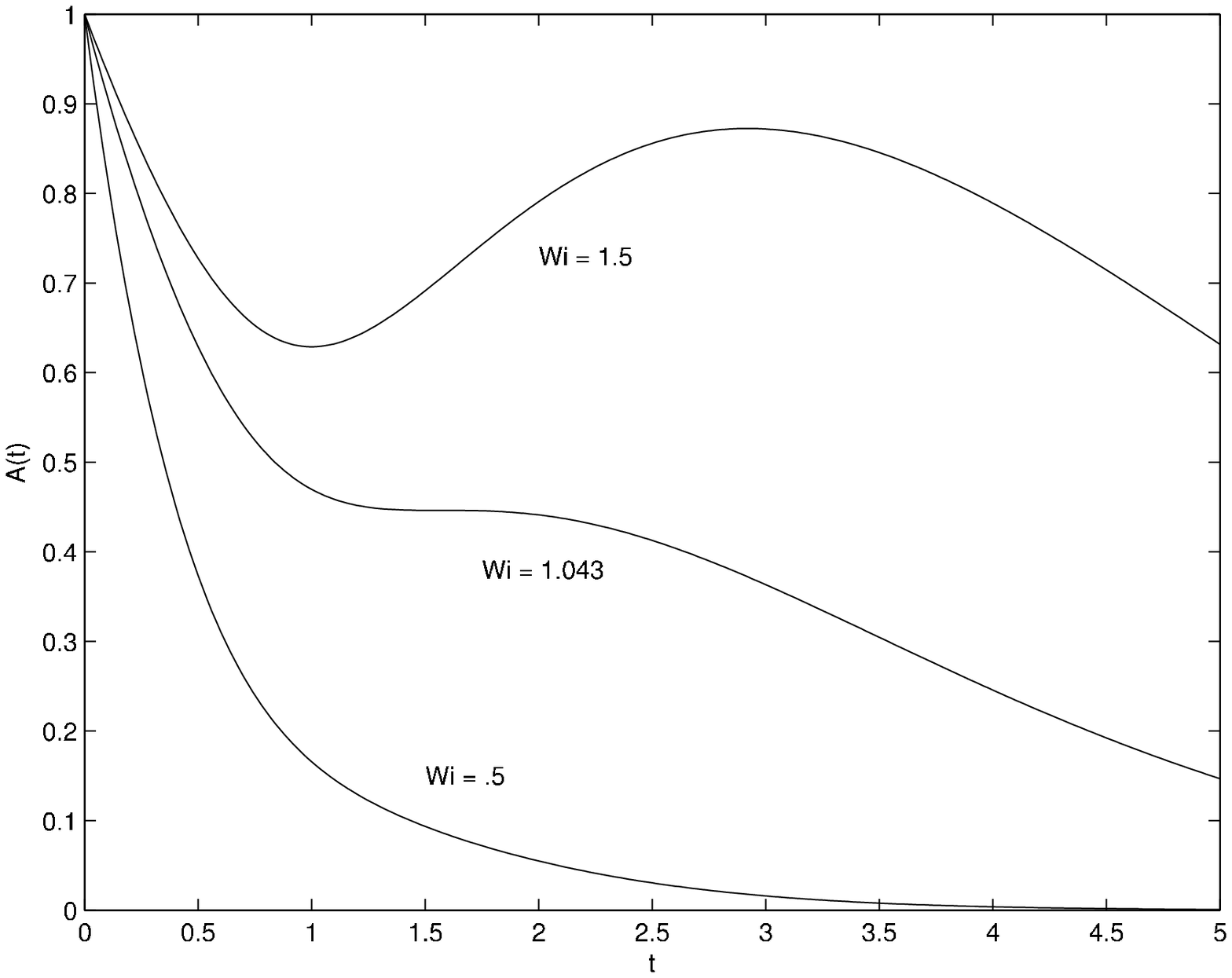}}
\caption[]{Transient growth of the perturbation amplitude $A(t)$ as a function of time
and Weissenberg number as given by Eq.~(\ref{amplitude}).} 
\label{pdf1}
\end{figure}

\clearpage

\begin{figure}
\centerline{\includegraphics[angle=0,scale=0.5,draft=false]{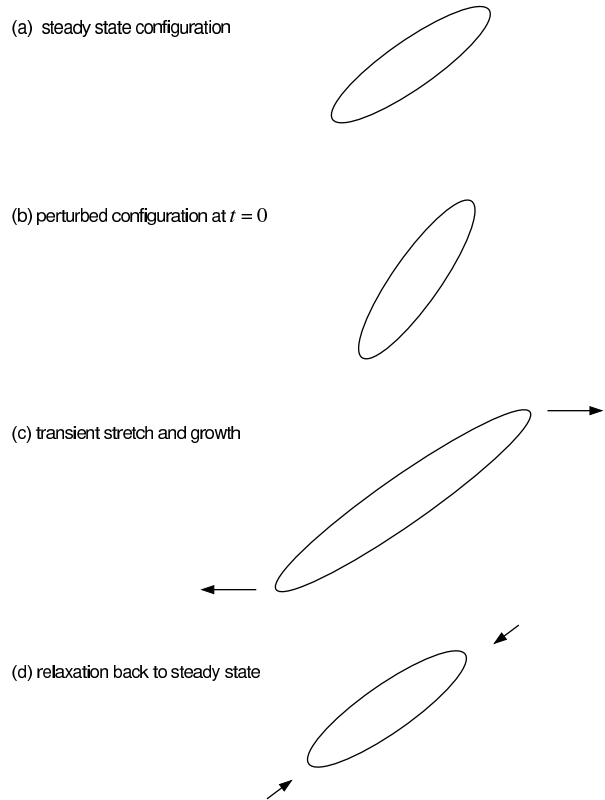}}
\caption[]{Sketch of the sequence of deformations of a polymer chain during the non-normal
amplification phase in a shear flow.} 
\label{pdf2}
\end{figure}

\end{document}